\documentclass[twocolumn]{aastex631}

\usepackage{graphicx}
\usepackage{amsmath}
\usepackage[utf8x]{inputenc}
\usepackage[encapsulated]{CJK}
\usepackage{ucs}
\usepackage{chngcntr}
\graphicspath{{./}{figs/}}

\newcommand{\cue}{\texttt{Cue}}

\newcommand{\prospector}{\texttt{Prospector}}
\newcommand{\cloudy}{\texttt{Cloudy}}

\newcommand{\msun}{{\rm M}_{\odot}}

\newcommand{\zphot}{z_{\rm phot}}
\newcommand{\zspec}{z_{\rm spec}}

\shorttitle{Systematics from Model Choices in SED Fitting}
\shortauthors{Wang et al.}

\begin{document}

\title{Quantifying the Effects of Known Unknowns on Inferred High-redshift Galaxy Properties: Burstiness, the IMF, and Nebular Physics}

\correspondingauthor{Bingjie Wang}
\email{bwang@psu.edu}

\author[0000-0001-9269-5046]{Bingjie Wang (\begin{CJK*}{UTF8}{gbsn}王冰洁\ignorespacesafterend\end{CJK*})}
\affiliation{Department of Astronomy \& Astrophysics, The Pennsylvania State University, University Park, PA 16802, USA}
\affiliation{Institute for Computational \& Data Sciences, The Pennsylvania State University, University Park, PA 16802, USA}
\affiliation{Institute for Gravitation and the Cosmos, The Pennsylvania State University, University Park, PA 16802, USA}

\author[0000-0001-6755-1315]{Joel Leja}
\affiliation{Department of Astronomy \& Astrophysics, The Pennsylvania State University, University Park, PA 16802, USA}
\affiliation{Institute for Computational \& Data Sciences, The Pennsylvania State University, University Park, PA 16802, USA}
\affiliation{Institute for Gravitation and the Cosmos, The Pennsylvania State University, University Park, PA 16802, USA}

\author[0000-0002-7570-0824]{Hakim Atek}
\affiliation{Institut d'Astrophysique de Paris, CNRS, Sorbonne Universit\'e, 98bis Boulevard Arago, 75014, Paris, France}

\author[0000-0002-2057-5376]{Ivo Labb\'e}
\affiliation{Centre for Astrophysics and Supercomputing, Swinburne University of Technology, Melbourne, VIC 3122, Australia}

\author[0000-0002-0682-3310]{Yijia Li (\begin{CJK*}{UTF8}{gbsn}李轶佳\ignorespacesafterend\end{CJK*})}
\affiliation{Department of Astronomy \& Astrophysics, The Pennsylvania State University, University Park, PA 16802, USA}
\affiliation{Institute for Gravitation and the Cosmos, The Pennsylvania State University, University Park, PA 16802, USA}

\author[0000-0001-5063-8254]{Rachel Bezanson}
\affiliation{Department of Physics \& Astronomy and PITT PACC, University of Pittsburgh, Pittsburgh, PA 15260, USA}
\author[0000-0003-2680-005X]{Gabriel Brammer}
\affiliation{Cosmic Dawn Center (DAWN), Niels Bohr Institute, University of Copenhagen, Jagtvej 128, K{\o}benhavn N, DK-2200, Denmark}
\author[0000-0002-7031-2865]{Sam E. Cutler}
\affiliation{Department of Astronomy, University of Massachusetts, Amherst, MA 01003, USA}
\author[0000-0001-8460-1564]{Pratika Dayal}
\affiliation{Kapteyn Astronomical Institute, University of Groningen, 9700 AV Groningen, The Netherlands}
\author[0000-0001-6278-032X]{Lukas J. Furtak}
\affiliation{Department of Physics, Ben-Gurion University of the Negev, P.O. Box 653, Be'er-Sheva 84105, Israel}
\author[0000-0002-5612-3427]{Jenny E. Greene}
\affiliation{Department of Astrophysical Sciences, Princeton University, Princeton, NJ 08544, USA}
\author[0000-0002-5588-9156]{Vasily Kokorev}
\affiliation{Kapteyn Astronomical Institute, University of Groningen, 9700 AV Groningen, The Netherlands}
\author[0000-0002-9651-5716]{Richard Pan}
\affiliation{Department of Physics \& Astronomy, Tufts University, MA 02155, USA}
\author[0000-0002-0108-4176]{Sedona H. Price}
\affiliation{Department of Physics \& Astronomy and PITT PACC, University of Pittsburgh, Pittsburgh, PA 15260, USA}
\author[0000-0002-1714-1905]{Katherine A. Suess}
\altaffiliation{NHFP Hubble Fellow}
\affiliation{Kavli Institute for Particle Astrophysics and Cosmology and Department of Physics, Stanford University, Stanford, CA 94305, USA}
\author[0000-0003-1614-196X]{John R. Weaver}
\affiliation{Department of Astronomy, University of Massachusetts, Amherst, MA 01003, USA}
\author[0000-0001-7160-3632]{Katherine E. Whitaker}
\affiliation{Department of Astronomy, University of Massachusetts, Amherst, MA 01003, USA}
\affiliation{Cosmic Dawn Center (DAWN), Niels Bohr Institute, University of Copenhagen, Jagtvej 128, K{\o}benhavn N, DK-2200, Denmark}
\author[0000-0003-2919-7495]{Christina C. Williams}
\affiliation{NSF's National Optical-Infrared Astronomy Research Laboratory, Tucson, AZ 85719, USA}
\affiliation{Steward Observatory, University of Arizona, Tucson, AZ 85721, USA}

\begin{abstract}

The era of the James Webb Space Telescope ushers stellar population models into uncharted territories, particularly at the high-redshift frontier.
In a companion paper, we apply the \texttt{Prospector} Bayesian framework to jointly infer galaxy redshifts and stellar population properties from broad-band photometry as part of the UNCOVER survey. 
Here we present a comprehensive error budget in spectral energy distribution (SED) modeling.
Using a sample selected to have photometric redshifts higher than 9, we quantify the systematic shifts stemming from various model choices in inferred stellar mass, star formation rate (SFR), and age. These choices encompass different timescales for changes in the star formation history (SFH), non-universal stellar initial mass functions (IMF), and the inclusion of variable nebular abundances, gas density and ionizing photon budget.
We find that the IMF exerts the strongest influence on the inferred properties: the systematic uncertainties can be as much as 1 dex, 2--5 times larger than the formal reported uncertainties in mass and SFR; and importantly, exceed the scatter seen when using different SED fitting codes. 
Although the assumptions on the lower end of the IMF induce degeneracy, our findings suggest that a common practice in the literature of assessing uncertainties in SED-fitting processes by comparing multiple codes is substantively underestimating the true systematic uncertainty.
Highly stochastic SFHs change the inferred SFH by much larger than the formal uncertainties, and introduce $\sim 0.8$ dex systematics in SFR averaged over short time scale and $\sim 0.3$ dex systematics in average age.
Finally, employing a flexible nebular emission model causes $\sim 0.2$ dex systematic increase in mass and SFR, comparable to the formal uncertainty.
This paper constitutes an initial step toward a complete uncertainty estimate in SED modeling.

\end{abstract}


\section{Introduction}

Since the initial data releases from the James Webb Space Telescope (JWST), rapid progress has been made in photometrically identifying $z>9$ candidates, leading to a new census of the early stages of galaxy formation (e.g., \citealt{Finkelstein2022,Naidu2022,Donnan2023,Harikane2023}). Among the surprising findings is the apparent over-abundance of high-redshift sources in comparison to theoretical predictions, or extrapolation of lower-redshift luminosity functions \citep{Bouwens2023,Castellano2023,Ferrara2023,Mason2023,Mauerhofer2023,Perez-Gonzalez2023}. Spectroscopic follow-ups of some of the candidates generally confirm their high-redshift nature, but also reveal lower-redshift interlopers \citep{2023arXiv230315431A,Curtis-lake2023,Wang2023:z12}.

Two questions immediately emerge from the early searches for $z>9$ candidates: the reliability of photometric redshifts and uncertainties in their key properties as inferred by stellar population synthesis (SPS) models.
The flexibility in SPS modeling enabled by Bayesian SED fitting codes \citep{Chevallard2016,Carnall2018,Johnson2021} means that these systematics can be investigated in detail.
In a companion paper \citep{Wang2023:uncover}, we release an SPS catalog, consisting of $\sim$~50,000 redshifts and stellar population properties for sources in the Abell~2744 galaxy cluster field, as part of the UNCOVER survey \citep{Bezanson2022}. The UNCOVER SPS catalog is a large-scale application of the \prospector\ Bayesian framework \citep{Johnson2021}, and is representative of the state-of-the-art in SPS modeling. Importantly, it provides a baseline for evaluating alternative SPS model choices in an exotic early universe now made eminently accessible by JWST.

SPS models include multiple ingredients, such as an initial mass function (IMF), stellar isochrones, and stellar spectra for the construction of simple stellar populations, SFH models and stellar metallicity models for the construction of composite stellar populations, a model for dust attenuation and emission, and nebular continuum and line emission (see \citealt{Conroy2013} for a review, and Section~5.2 of \citealt{Wang2023:uncover} for a discussion in the context of the UNCOVER SPS catalog). Even with data of exquisite quality and wavelength coverage at $z \lesssim 3$, the dependence of inferred parameters on modeling assumptions is already visible \citep{Pacifici2023}. At $z \gtrsim 10$, SPS models have yet to undergo robust tests. This implies that the dominant uncertainty for high-redshift measurements is not usually measurement uncertainty, but instead systematic uncertainty which can potentially be orders of magnitudes higher than the former. 
The implications of different stellar libraries and functional forms of SFHs for JWST observations have been studied in some length \citep{Whitler2023}. This work focuses on a different set of systematics, encompassing the following three model choices.

First, a non-universal IMF. The relative numbers of stars as a function of the stellar mass influence most of the observable properties of stellar populations. The strongest effect is expected on the inferred stellar mass: at rest-frame optical/ultraviolet (UV) wavelengths for rising SFHs, we only observe $M \gtrsim 20 \msun$ stars (e.g., Figure~5 in \citealt{Conroy2013}), meaning that a considerable fraction of the reported stellar mass relies on an extrapolation from the assumed IMF shape.
Despite such importance, a complete theoretical understanding of the origin of the shape of the IMF is still lacking (see \citealt{Bastian2010,Smith2020} for recent reviews). Introduced in \citet{Salpeter1955}, the functional form of the IMF above the solar mass scale remains remarkably consistent with a slope of $\sim-1.3$. Since then, follow-up studies of the Galactic regions suggest deviations from the nominal IMF \citep{Kroupa2001,Chabrier2003}, but it is typically assumed to be constant cross time and environment. There is, however, mounting evidence in the nearby universe against a universal IMF \citep{Treu2010,Conroy2012,Geha2013,McWilliam2013,vanDokkum2017}. Recent studies have claimed that IMF variations may be particularly relevant in the context of early star-forming activities \citep{Chon2022,Katz2022,Pacucci2022,Steinhardt2023}. In this work we aim to quantify the systematic effects of IMFs that cover a reasonable range as suggested by observational studies.

Second, outshining. It has long been known that bursty star formation causes light from recent star formation to outshine that of older stellar populations \citep{Papovich2001}. Here burstiness refers to the pattern of sporadic star-forming activities interspersed with less active or ``quiescent'' phases. Such stochasticity is expected at high redshift as the feedback timescale approaches or exceeds the dynamical time of the system, decreasing and delaying the regulatory effect on the star formation rate (SFR) \citep{Tacchella2016,Faucher-Gigueere2018,Dome2023}. Observations also hint at bursty SFHs \citep{Caputi2007,Smit2015,Guo2016,Looser2023}.
A recent bursty star formation in the detected high-redshift galaxies is also strongly motivated from their extremely blue UV slopes, which are explained by ejection of dust due to sudden enhancement of radiation pressure \citep{Tsuna2023,Ziparo2023}.
 The recent SFH significantly influences galaxy SEDs across the electromagnetic spectrum and particularly in the rest-frame UV, meaning that the inference of galaxy properties depends on the assumed level of burstiness \citep{Furlanetto2022,Endsley2023}. Various previous studies have investigated the effect of varying SFHs and priors \citep{Carnall2018,Leja2019:nonpar,Suess2022,Duan2023}. 
We focus on assumed burstiness in the context of nonparametric SFHs, and test it by varying the prior on the change in SFRs in adjacent age bins. Our ``bursty'' prior represents a more extreme case than what is implied by the bursty continuity prior in \citet{Tacchella2022:metal}. This is an intentional choice as the goal is to perform a sensitivity test rather than propose a new physical model.

Third, nebular physics. Line and continuum emission can make up a significant fraction of the total flux for stars at low metallicity and at young ages, up to 50\% of the flux in broadband filters for highly star-forming galaxies \citep{Pacifici2015}, implying increased importance at high redshift \citep{Charlot2001,Anders2003,Reines2010,Smit2014}. The most sophisticated approach of incorporating nebular physics into SED fitting thus far relies on a model grid pre-computed using photoionization simulations \citep{Gutkin2016,Byler2017}. While such an approach better accounts for variations in the recent SFH and non-solar metallicity than analytic prescriptions \citep{Fioc1999,Leitherer1999,Molla2009}, the accessible parameter space is still limited. 
The diverse chemical environment that may be present at early epochs (e.g., \citealt{Bunker2023,Katz2023}), and the possibility of strong optical nebular emission lines at $z < 6$ mimicking JWST/NIRCam photometry of high-redshift galaxies \citep{McKinney2023,Zavala2023} renew the interest in the modeling of nebular emission with increased flexibility in elemental abundance ratios and separate parameterizations of gas-phase number density and ionizing photon density, while retaining consistency with the ionizing flux emitted by the model stellar populations.
We do so by integrating a neural net emulator for the spectral synthetic code \cloudy\ \citep{Chatzikos2023}, dubbed \cue\ (Li et al. in prep.), into \prospector.

The structure of this paper is as follows. Section~\ref{sec:data} summarizes the JWST observations, including our sample of $z > 9$ candidates. Section~\ref{sec:new_fit} details our fiducial as well as new modeling approaches regarding the IMF, SFH prior, and nebular physics. Section~\ref{sec:res} presents the effects of new models on the inferred parameters. We discuss the error budget as well as possible ways forward in Section~\ref{sec:discussion}, and conclude in Section~\ref{sec:conclusion}.

Where applicable, we adopt the best-fit cosmological parameters from the WMAP 9 yr results: $H_{0}=69.32$ ${\rm km \,s^{-1} \,Mpc^{-1}}$, $\Omega_{M}=0.2865$, and $\Omega_{\Lambda}=0.7135$ \citep{Hinshaw2013}. Unless otherwise mentioned, we report the median of the posterior, and 1$\sigma$ error bars are the 16th and 84th percentiles.

\begin{deluxetable*}{lccllll}
\tablecaption{$\zphot>9$ Candidates Studied in this Paper \label{tab:mysample}}
\tablehead{
\colhead{ID} & \colhead{RA} & \colhead{Dec} & \colhead{$z_{\rm phot}$\tablenotemark{\scriptsize{a}}} & \colhead{$M_\star$\tablenotemark{\scriptsize{a}}} & $\zspec$\tablenotemark{\scriptsize{b}} & \colhead{Other IDs\tablenotemark{\scriptsize{c}}}\\
\colhead{} & \colhead{[J2000 deg]} & \colhead{[J2000 deg]} & \colhead{} & \colhead{[log~$\msun$]} & \colhead{} & \colhead{} 
}
\startdata
4745 & 3.61720 & -30.42554 & $9.957^{+0.290}_{-0.301}$ & $9.215^{+0.219}_{-0.150}$ & 9.325 (3686; F23) & dhz1 / 2065 \\
7720 & 3.60383 & -30.41581 & $9.157^{+0.149}_{-0.250}$ & $7.285^{+0.231}_{-0.193}$ & & \\
14019 & 3.57087 & -30.40158 & $13.631^{+0.591}_{-0.466}$ & $7.866^{+0.250}_{-0.170}$ & 13.079 (UNCOVER-z13; W23) & \\
14088 & 3.59250 & -30.40146 & $10.104^{+0.097}_{-0.124}$ & $7.526^{+0.295}_{-0.192}$ & 9.880 (13151; F23) & \\
27025 & 3.56707 & -30.37786 & $10.975^{+0.226}_{-0.331}$ & $7.834^{+0.235}_{-0.140}$ & 10.071 (26185; F23) & uhz1 / 21623  \\
28397 & 3.56152 & -30.37494 & $10.098^{+0.201}_{-0.163}$ & $7.059^{+0.149}_{-0.133}$ &   & \\
30818 & 3.51193 & -30.37186 & $9.868^{+0.750}_{-0.353}$ & $8.881^{+0.366}_{-0.235}$ & & ghz1 / 26928 \\
38095 & 3.59011 & -30.35974 & $10.928^{+0.250}_{-0.261}$ & $8.163^{+0.172}_{-0.150}$ & 10.255 (37126; F23) & 39074 \\
39753 & 3.51356 & -30.35680 & $13.474^{+0.816}_{-0.689}$ & $8.047^{+0.267}_{-0.205}$ & 12.393 (UNCOVER-z12; W23) & \\
42485 & 3.51374 & -30.35157 & $10.272^{+0.521}_{-0.608}$ & $8.391^{+0.224}_{-0.221}$ &   & ghz4  \\
42658\tablenotemark{\scriptsize{d}} & 3.52286 & -30.35129 & $18.915^{+0.432}_{-0.445}$ & $8.207^{+0.181}_{-0.158}$ & & \\
45758 & 3.47874 & -30.34554 & $11.067^{+0.248}_{-0.257}$ & $8.477^{+0.222}_{-0.185}$ &   & ghz9 / 52008 \\
46843 & 3.50731 & -30.34321 & $11.119^{+0.398}_{-0.450}$ & $7.898^{+0.216}_{-0.174}$ &   &  \\
54328 & 3.49899 & -30.32476 & $12.559^{+0.043}_{-0.054}$ & $8.673^{+0.158}_{-0.161}$ & 12.117 (GLASS-z12; B23) & ghz2 / 70846 \\
55541 & 3.45142 & -30.32181 & $11.341^{+0.253}_{-0.283}$ & $8.412^{+0.187}_{-0.130}$ &   & ghz8 / 73667 \\
55950 & 3.45136 & -30.32072 & $10.802^{+0.310}_{-0.869}$ & $8.307^{+0.219}_{-0.154}$ &   & ghz7 / 81198 \\
57149 & 3.45471 & -30.31689 & $9.985^{+0.462}_{-0.452}$ & $9.043^{+0.321}_{-0.280}$ & & 83338 \\
61062 & 3.49177 & -30.29831 & $9.118^{+0.124}_{-0.149}$ & $8.461^{+0.165}_{-0.160}$ & & \\
\enddata
\tablenotetext{a}{Posterior moments of photometric redshifts and stellar masses from \prospector\ fits.}
\tablenotetext{b}{Spectroscopic redshifts. All are determined from NIRSpec/Prism data, except GLASS-z12, which is based on an [O \textsc{iii}] detection from ALMA \citep{Bakx2023}. F23 refers to \citet{Fujimoto2023:uncover}, and W23 refers to \citet{Wang2023:z12}.}
\tablenotetext{c}{Photometrically identified candidates in the literature. Alphabetical IDs are from \citet{Castellano2022}, and numerical IDs are from \citet{Atek2023:uncover}; GLASS-z12 is also selected in \citet{Naidu2022}.}
\tablenotetext{d}{Possible data quality issue; see the Appendix for detail.}
\end{deluxetable*}

\section{Data\label{sec:data}}

\subsection{Photometry}

This paper utilizes the latest photometric catalog \citep{Weaver2023}, made available as part of the second data release from the UNCOVER survey \citep{Bezanson2022}\footnote{All data products are accessible via \url{https://jwst-uncover.github.io/\#releases}}. The catalog includes all publicly available imaging data of Abell 2744 obtained from JWST/NIRCam, HST/ACS, and HST/WFC3. In particular, the JWST observations consist of three programs: UNCOVER (PIs Labb\'e \& Bezanson, JWST-GO-2561; \citealt{Bezanson2022}), the Early Release Science program GLASS (PI: Treu, JWST-ERS-1324; \citealt{Treu2022}), and a Director's Discretionary program (PI: Chen, JWST-DD-2756). These observations cover a wavelength range of $\sim 1-5 \mu$m in the observed frame, using 8 filters: F090W, F115W, F150W, F200W, F277W, F356W, F410M, and F444W. The HST data, obtained from the public archive, include HST-GO-11689 (PI: Dupke), HST-GO-13386 (PI: Rodney), HST-DD-13495 (PI: Lotz; \citealt{Lotz2017}), and HST-GO-15117 (PI: Steinhardt; \citealt{Steinhardt2020}). These additional observations cover a wavelength range of $\sim 0.4-1.6 \mu$m in the observed frame, using 7 filters: F435W, F606W, F814W, F105W, F125W, F140W, and F160W.

\subsection{Fiducial Redshifts and Stellar Population Properties\label{sec:fit}}

Fiducial redshifts and galaxy properties are taken from the UNCOVER SPS catalog \citep{Wang2023:uncover}. Here we only briefly reiterate the modeling components for completeness. All parameters, including redshift, are inferred jointly using the \prospector\ inference framework \citep{Johnson2021}, adopting the MIST stellar isochrones \citep{Choi2016,Dotter2016} and MILES stellar library \citep{Sanchez-Blazquez2006} from FSPS \citep{Conroy2010}. 
A non-parametric form is assumed for the SFH, defined by mass formed in 7 logarithmically-spaced time bins (\prospector-$\alpha$; \citealt{Leja2017}). 
Only the contribution of obscured active galactic nuclei (AGN) is considered following \citet{Nenkova2008a,Nenkova2008b}, and parameterized by the normalization and dust optical depth in the mid-infrared \citep{Conroy2009,Leja2018}.
A mass function prior, and a dynamic SFH($M, z$) prior are included to optimize the photometric inference over the wide parameter space covered by deep JWST surveys \citep{Wang2023:pbeta}. Lensing magnification, computed from the public \texttt{v1.1} UNCOVER lensing maps by \citet{Furtak2023}, is performed on-the-fly to ensure consistency with the scale-dependent priors.
The model consists of 18 free parameters, and sampling is performed using the dynamic nested sampler \texttt{dynesty} \citep{Speagle2020} with a speed up in model generation enabled by a neural net emulator dubbed \texttt{parrot} \citep{Mathews2023}.

\subsection{High-redshift Candidates\label{subsec:sample}}

We select $\zphot > 9$ candidates directly from the newly-public SPS catalog, demonstrating the quality of the catalog as well as the exciting parameter space now accessible via deep JWST surveys. Specifically, the criteria are

\begin{enumerate}
\item Redshift posteriors in the \prospector-$\beta$ \citep{Wang2023:pbeta} fits
\begin{enumerate}
   \item 50th percentile of the redshift posteriors $\geq 9.0$
   \item 0.1th percentile of the redshift posteriors $\geq 3.0$
 \end{enumerate}
\item Maximum-likelihood redshift $\geq 8.0$
\item $\chi^2_{\rm red} = \chi^2/n_{\rm bands} < 5$
\item Number of NIRCam filters $> 6$
\item S/N in photometry 
\begin{enumerate}
   \item F444W $\geq 2.0$
   \item F410M $\geq 2.0$ if available
   \item F277W $\geq 6.0$
 \end{enumerate}
 \end{enumerate}
 
The first criterion defines our selection---criterion 1-a imposes the redshift cut, whereas 1-b decreases the probability of low-redshift interpolators given the multi-modality commonly seen in the posterior distribution. The rest of the criteria is to ensure the purity of the sample: $\chi^2_{\rm red} < 5$ is applied to ensure a reasonable fit; that is, to remove data defects in the photometric catalog that are difficult to be described by SPS models. The S/N cut in the photometry is determined empirically after visual inspection of the 31 candidates given the first 4 criteria.

Our selection eventually yields 18 candidates, which are listed in Table~\ref{tab:mysample}. In the same table, we cross-match our sample to existing candidates in the literature, and include available spectroscopic redshifts as well. 
We note that the spectroscopic redshifts are systematically lower than the photometric redshifts by $\sim 1.5-2 \sigma$, which may be a manifestation of the ``Eddington bias" claimed in \citet{Serjeant2023}.
Further details on this sample are supplemented in the Appendix, which include the new candidates from this paper, a possible artifact, and discrepant photometric and spectroscopic redshifts.

In principle, the cutoff based on the maximum-likelihood redshifts is unnecessary if the posterior mass is located correctly and consistently in the full sample. However, as it is still in early stage of photometry calibration for JWST, we impose this criterion for a cleaner sample. In the future, removing this would provide a more independent approach for selecting high-redshift candidates, complementing the traditional color selection technique (e.g., \citealt{Castellano2022,Atek2023:color}). 

We now proceed to investigate the effects of various modeling choices on the inferred redshifts and galaxy properties.
It is worth noting that in order to reach a definitive conclusion regarding the signals of interest of this paper, relatively high S/N spectra with medium to high-resolution would likely be necessary. For instance, rest-frame near-infrared spectra with S/N $\gtrsim$ 100 are needed for placing constraints on the IMF in the local universe \citep{Conroy2012,vanDokkum2012}. Given the difficulty in acquiring such data at high redshift, the scope of this paper is restricted to exploring the systematic effects on inferred properties when fitting broad-band photometry.

\section{Varying Known Unknowns in Stellar Population Inference\label{sec:new_fit}}

\begin{figure*}
 \gridline{\fig{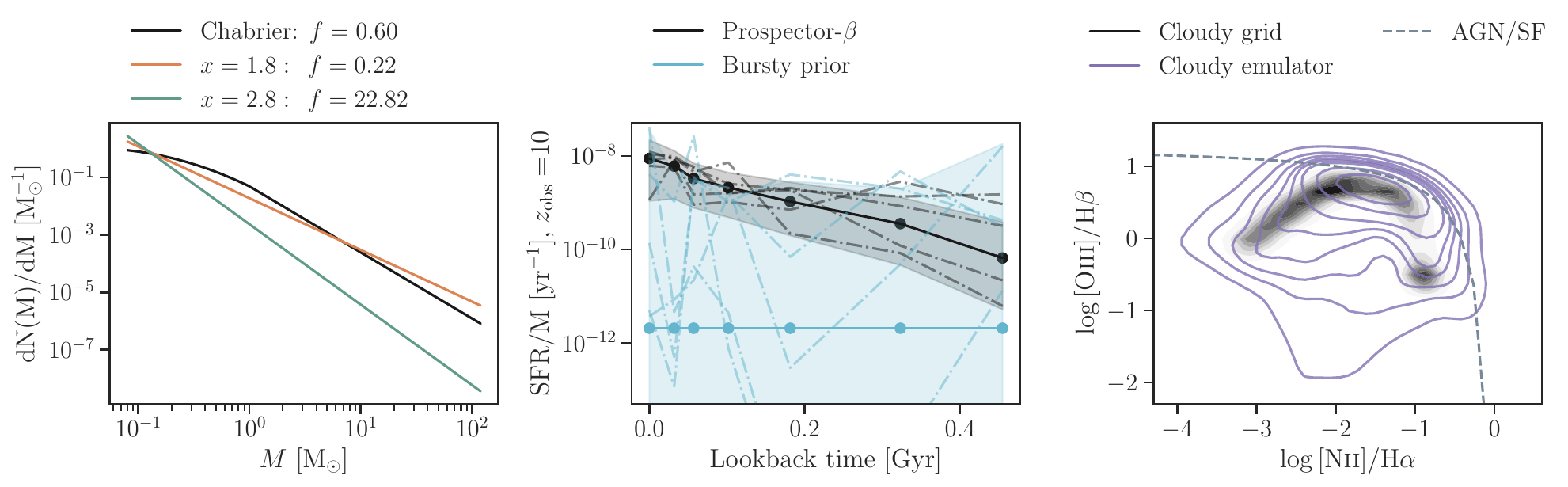}{0.99\textwidth}{}}
\caption{Model choices tested in this paper. (Left) The fiducial Chabrier IMF is plotted in black.
A flat ($x=1.8$) IMF is shown in orange, whereas a steep ($x=2.8$) IMF is shown in green.
$f$ is the ratio of the numbers of low-mass ($0.08< \msun{} < 0.5$) to massive ($5< \msun{} < 120$) stars.
(Middle) Prior distributions of specific SFRs for a galaxy observed at $z=10$ are plotted as functions of lookback time in Gyr. The fiducial rising SFH prior \citep{Wang2023:pbeta} is illustrated by the black curves, while the extremely bursty prior is illustrated by the cyan curves. Shading indicates 16th and 84th quantiles.
(Right) The BPT diagram demonstrates the model coverage in the standard \cloudy\ grid \citep{Byler2017} and in the more flexible model enabled by \cue. The gray dashed curve separating AGNs and starburst galaxies is taken from \citet{Kauffmann2003}.}
\label{fig:models}
\end{figure*}

While the current inference framework is highly flexible, the known unknowns in the SPS modeling may induce systematic shifts in the inferred parameters comparable to or larger than the uncertainties in the fiducial model. The companion paper \citep{Wang2023:uncover} discusses some of the assumptions that are particularly relevant to the UNCOVER SPS catalog. This work further examines the influences of a steeper/flatter IMF, bursty star formation, and more flexible nebular emission modeling, as summarized in Figure~\ref{fig:models}. We elaborate on both fiducial and alternative models below.

\subsection{Initial Mass Function}

The fiducial IMF is taken from \citet{Chabrier2003}. 
We consider two alternative IMFs described by a power law of slope $x$ over the full mass range as
\begin{equation}
	\xi \propto m^{-x},
\end{equation}
Two variations are examined, $x=1.8$ and $x=2.8$, covering a reasonable range as suggested from observations \citep{Cappellari2022,Martin-Navarro2015,vanDokkum2017}. 
For the main analyses, we define all the IMFs over the interval $0.08 < M_{\star} < 120 \, \msun$ for a controlled experiment. However, the minimum ($m_l$) and maximum stellar mass cutoffs remain uncertain. The low mass end of the IMF is particularly difficult to constrain since these stars are largely invisible outside the Milky Way. We thus additionally consider two cases where $m_l = 0.5\msun$ and $m_l = 1\msun$ \citep{Kroupa2001,Chabrier2003}.

Since we intend to test the scenario in which the IMF varies across cosmic time, and changing the IMF affects the scale of the predicted spectrum, it is necessary to modify the mass prior in \citet{Wang2023:pbeta} accordingly. We find the normalization factor by minimizing the $\chi^2$ between spectra predicted by different IMFs, and simply scale the mass prior by this factor. Strictly speaking, IMF($z$) describes instantaneous SFR, but stellar mass is determined by effective IMF from all stars formed. We thus note that our modification on the mass prior serves only as an approximation to mitigate the bias that may be induced by the scale difference among different IMFs. The fact that we assume the same IMF over the entire history justifies this simplification.

Effectively, the same stellar mass function prior is applied to all fits. Therefore, our choice of the prior has no influence on the systematic effects to be quantified as these are the relative shifts in the inferred parameter with respect to the fiducial values. In addition, the mass prior is constructed by using the observed mass functions between $0.2 < z < 3$ from \citet{Leja2020:mf}; for $z > 3$, we adopt the nearest-neighbor solution, i.e., the $z=3$ mass functions, in the absence of reliable high-resolution rest-frame optical selected mass functions at $z > 3$. As mentioned in \citet{Wang2023:pbeta}, this choice allots a conservatively high probability for yet-to-be-discovered populations of high-mass, high-redshift galaxies.

\subsection{Timescales for Changes in SFH}

All SFHs are modeled non-parametrically via mass formed in 7 logarithmically-spaced time bins \citep{Leja2017}. In our fiducial setup, we use the dynamic SFH prior from \citet{Wang2023:pbeta}. The expectation values of the logarithmic SFR ratios in adjacent time bins are based on the observed cosmic SFR density in the empirical UniverseMachine model \citep{Behroozi2019}, favoring rising SFHs in the early universe and falling SFHs in the late universe, and adjusted as a function of mass to reflect downsizing \citep{Cowie1996,Thomas2005}. The distribution of the logarithmic SFR ratios is modeled as a student-t distribution.

An additional bursty SFH is generated by replacing the student-t distribution with a uniform prior spanning from -5 to 5. This means that the SFR is effectively uncorrelated between the time bins, and can jump up to 10 orders of magnitude in a single time step. Our prior is similar in concept to the bursty continuity prior in \citet{Tacchella2022:metal}, but represents a more extreme bursty scenario.

\subsection{Nebular Physics}

\prospector\ models the nebular emission self-consistently via a pre-computed \cloudy\ grid \citep{Byler2017}. The parameter space is limited for a practical reason: grid-based approaches necessarily suffer from the curse of dimensionality.

We replace the grid with a neural net emulator, dubbed \cue\ (Li et al. in prep.), which emulates the continuum and line emission from a single H~\textsc{ii} region based on the spectral synthetic code \cloudy\ (version 22.00; \citealt{Chatzikos2023}). \cue\ is tailored to flexibly model exotic chemistry and unusual ionizing properties of galaxies, and is well-suited for galaxies in the early universe. It models the ionizing input spectra as a flexible 4-part piecewise-continuous power-law, along with freedom in gas density, total ionizing photon budget, O/H, C/O, and N/O. While in principle this power law approximation allows us to infer the ionizing spectrum separately from the model stellar spectra, here we fit this piecewise power law to the stellar ionizing spectra from the model for each spectrum generated. With the exception of carbon and nitrogen, the element abundances are scaled with the oxygen abundance, based on solar abundances and dust depletion factor specified in \citet{Dopita2000}.

The newly gained flexibility, however, does not {\it a priori} forbid unlikely nebular abundances. Certain chemical ratios are preferred on galactic scales due to the production rate of various elements; for example, N/O as a function of oxygen abundance has a well-defined shape following from the secondary production of nitrogen.
We thus place a Gaussian prior on the N/O ratio against oxygen abundance based on the observed trend. The Gaussian mean and standard deviation are estimated from the data compiled by \citet{2004ApJS..153....9G}. In the low oxygen abundance regime where no data is available, we take values from the model grid in \citet{Gutkin2016}.

\section{Results\label{sec:res}}

\begin{figure*}
 \gridline{
 \fig{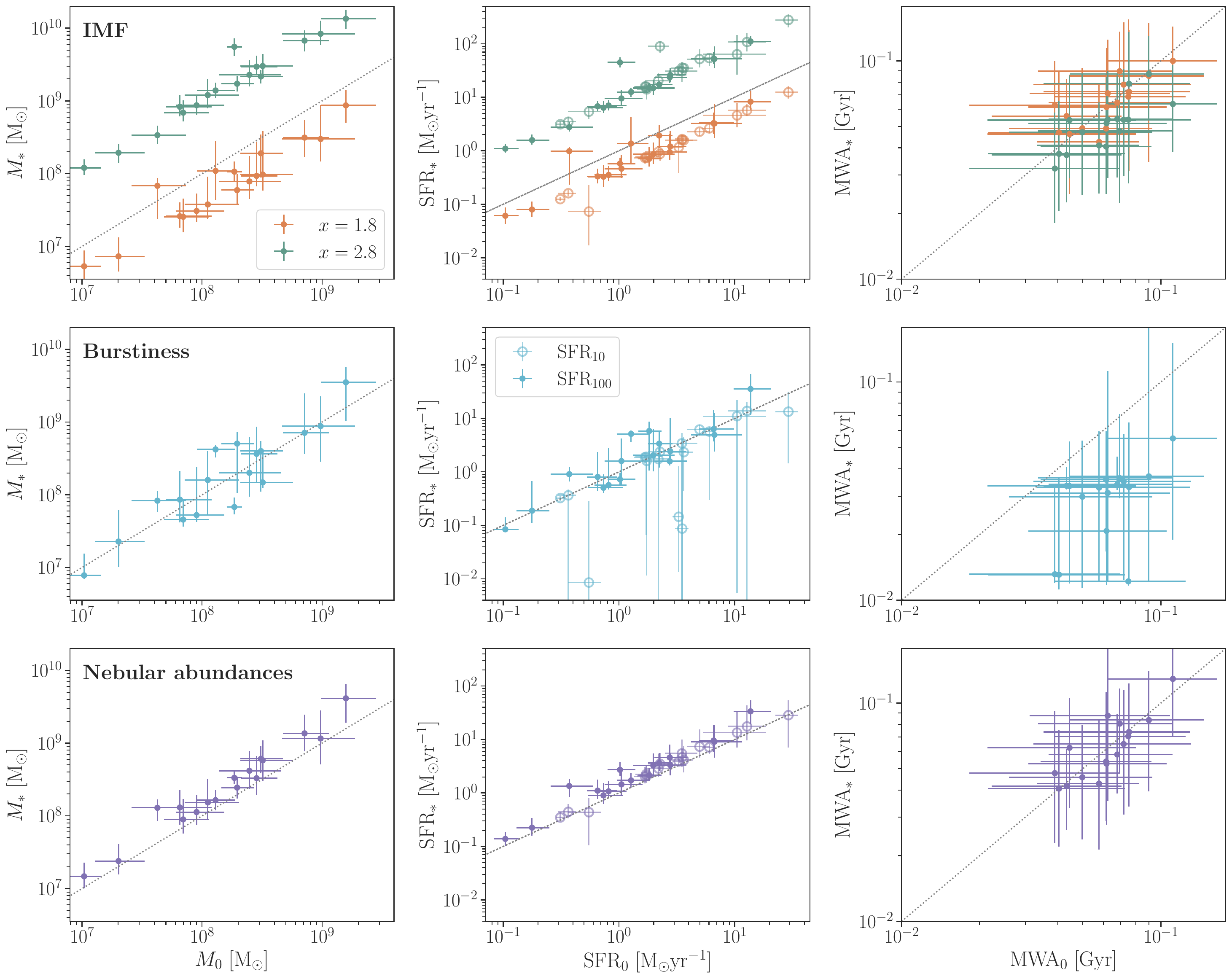}{0.99\textwidth}{}
 }
\caption{Changes in inferred parameter due to different model choices. The stellar mass, SFR, and mass-weighted age (MWA) are plotted against the fiducial values in each panel. (Upper panel) The comparison between results assuming an IMF with $x=1.8$ ($x=2.8$) and the fiducial Chabrier IMF are shown in orange (green). Significant systematic shifts in stellar mass and SFR are found. (Middle panel) The fiducial model assumes a rising SFH prior \citep{Wang2023:pbeta}, while the alternative model assumes a bursty prior. The SFRs averaged over the most recent 10 Myr (SFR$_{10}$, unfilled circles show significant scatter driven by the stochasticity in the SFH, but the SFRs averaged over the most recent 100 Myr (SFR$_{100}$, filled circles) are less affected. The a bursty prior also cause a systematic decrease in stellar age. (Lower panel) The fiducial model computes nebular emission and continuum from a pre-computed \cloudy\ grid \citep{Byler2017}, while the alternative model uses a \cloudy\ emulator dubbed \cue. The extra flexibility enabled by \cue\ systematically increases the inferred mass and SFR by $\sim 0.1-0.2$ dex, similar in size to the $1\sigma$ measurement uncertainties indicated by the error bars.}
\label{fig:scatter}
\end{figure*}

The effect of these changes on the stellar mass, SFR, and stellar age are summarized in Figures~\ref{fig:scatter}, where these key parameters inferred from alternative models are plotted against the fiducial values.
Furthermore, the preference of a given model by the data is quantified via the Bayesian evidence in Figure~\ref{fig:lnz}. 
The results regarding each model are presented in more detail below.

\begin{figure*}
 \gridline{
 \fig{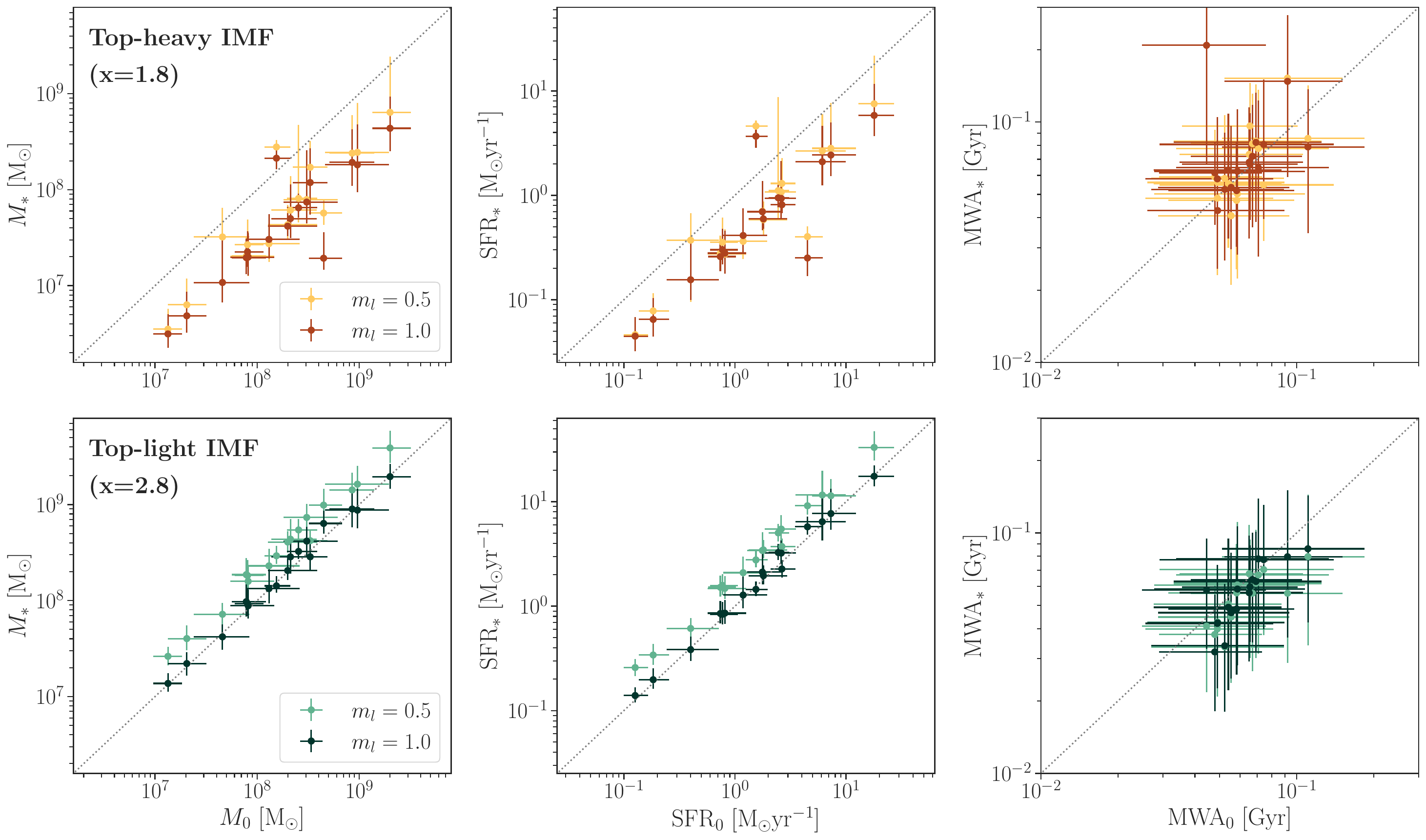}{0.99\textwidth}{}
 }
\caption{Same as Figure~\ref{fig:scatter}, but showing the changes in the inferred parameters with increasing minimum stellar mass cutoffs, $m_l$, of a given IMF.}
\label{fig:scatter_ml}
\end{figure*}

\begin{figure*}
 \gridline{\fig{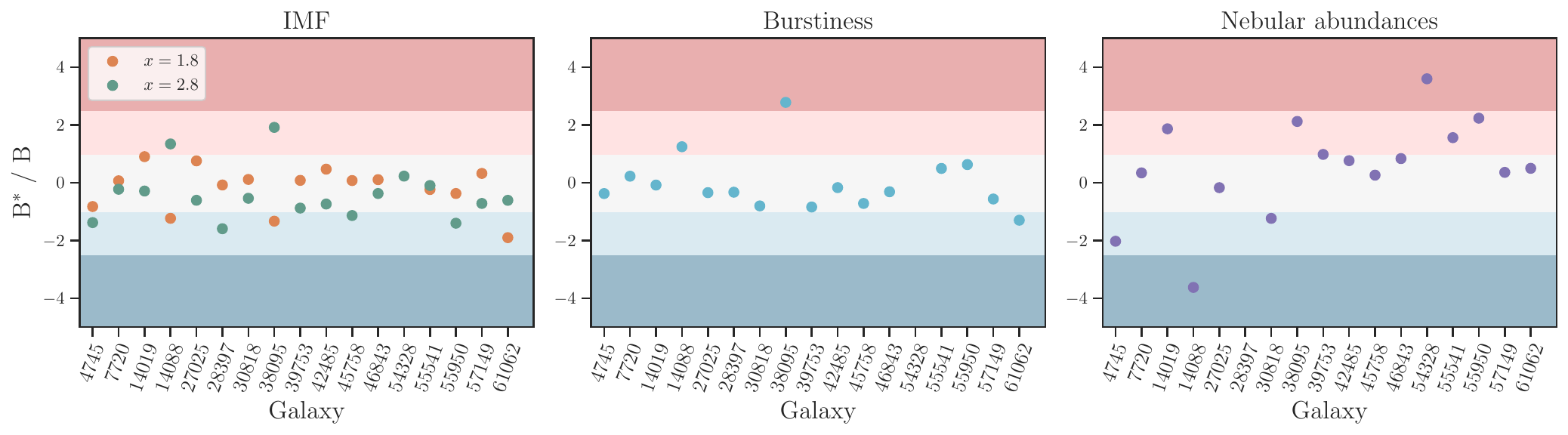}{0.99\textwidth}{}}
\caption{Model comparison. Logarithm of the Bayes factor comparing different model choices is calculated for each candidate. A value of $\rm B^*/B > (<)~1$ represents an increase (decrease) of the support in favor of the alternative model given the observed data \citep{Trotta2008}. The boundaries between different shading indicate weak evidence ($\rm |B^*/B| = 1$), moderate evidence ($\rm |B^*/B| = 2.5$), and strong evidence ($\rm |B^*/B| = 5$), respectively. In most cases, the data are agnostic of the model choices.}
\label{fig:lnz}
\end{figure*}

\begin{figure*}
 \gridline{
 \fig{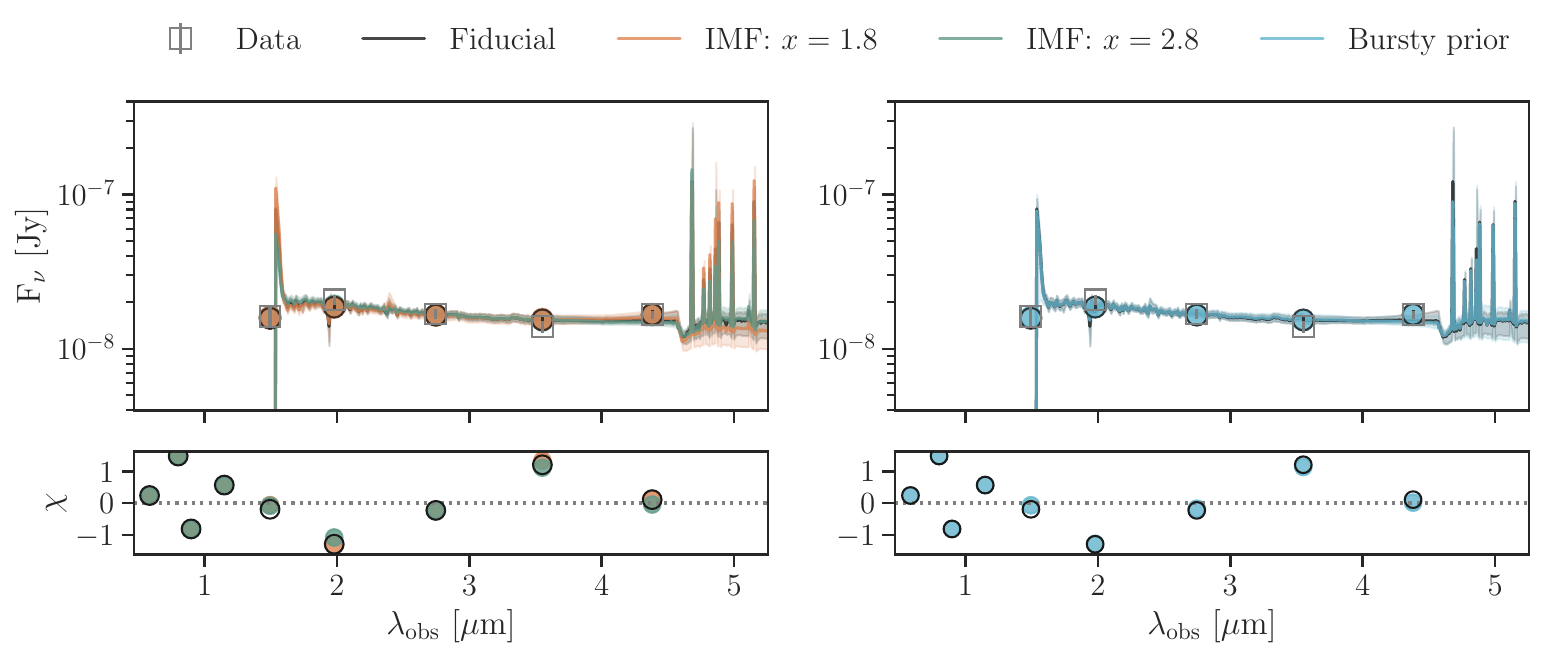}{0.99\textwidth}{(a)}
 }
 \gridline{
 \fig{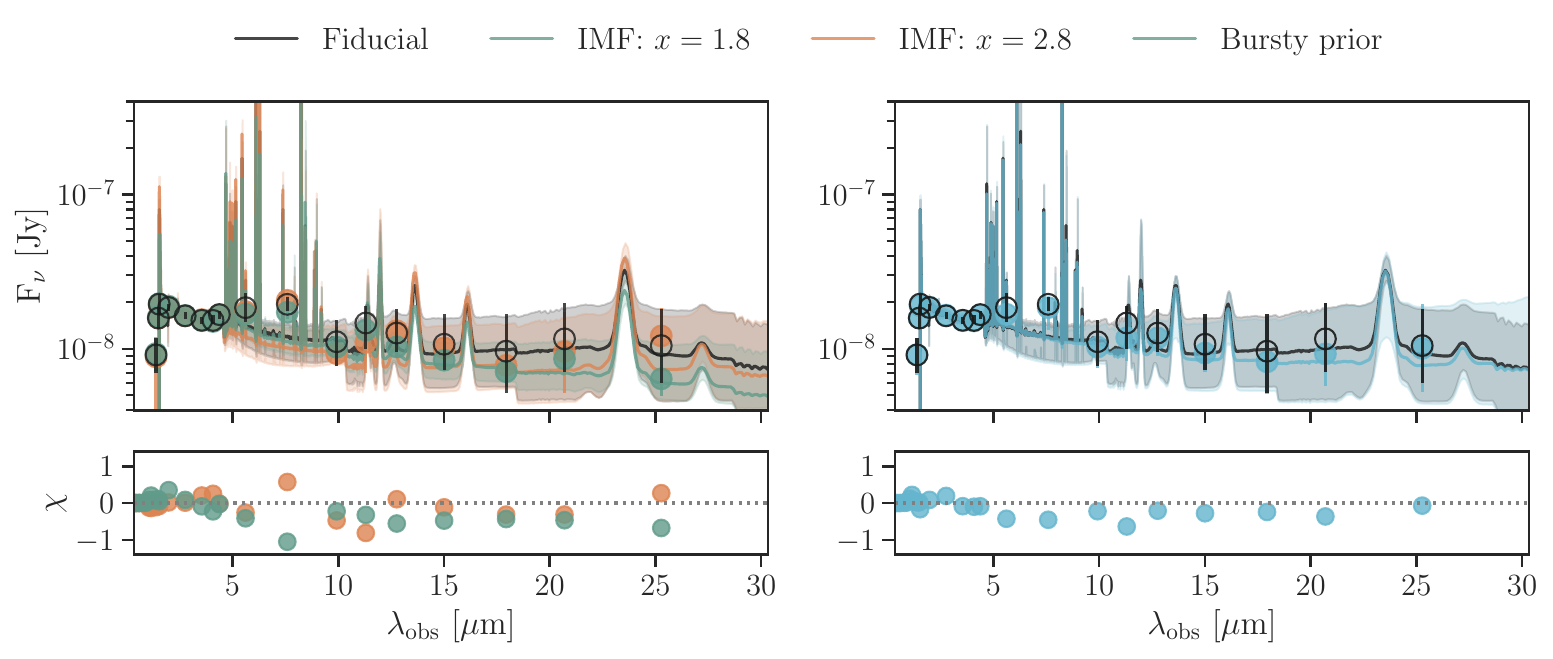}{0.99\textwidth}{(b)}
 }
 \caption{Model photometry and spectra for one of the galaxies as an example. (a) Photometric data are shown as gray squares. Median model spectra assuming the different IMFs are plotted as functions of observed wavelengths in the left panel, and the shading indicates the 16th and 84th quantiles. Median model photometries are over-plotted in the same colors. The right panel is the same, but compares the fiducial model with one which assumes a bursty prior. In all cases, the model photometric points fit the data equally well, and appear to be identical, whereas the model spectra exhibits marginal difference. (b) Same as the upper panel, but including the longer wavelength range covered by JWST/MIRI. The lower panels here show the differences between the photometries of the alternative models and the fiducial model, normalized by the 1$\sigma$ uncertainty in the fiducial model photometry.
 }
\label{fig:eg_modspec}
\end{figure*}

\begin{figure*}
 \gridline{
 \fig{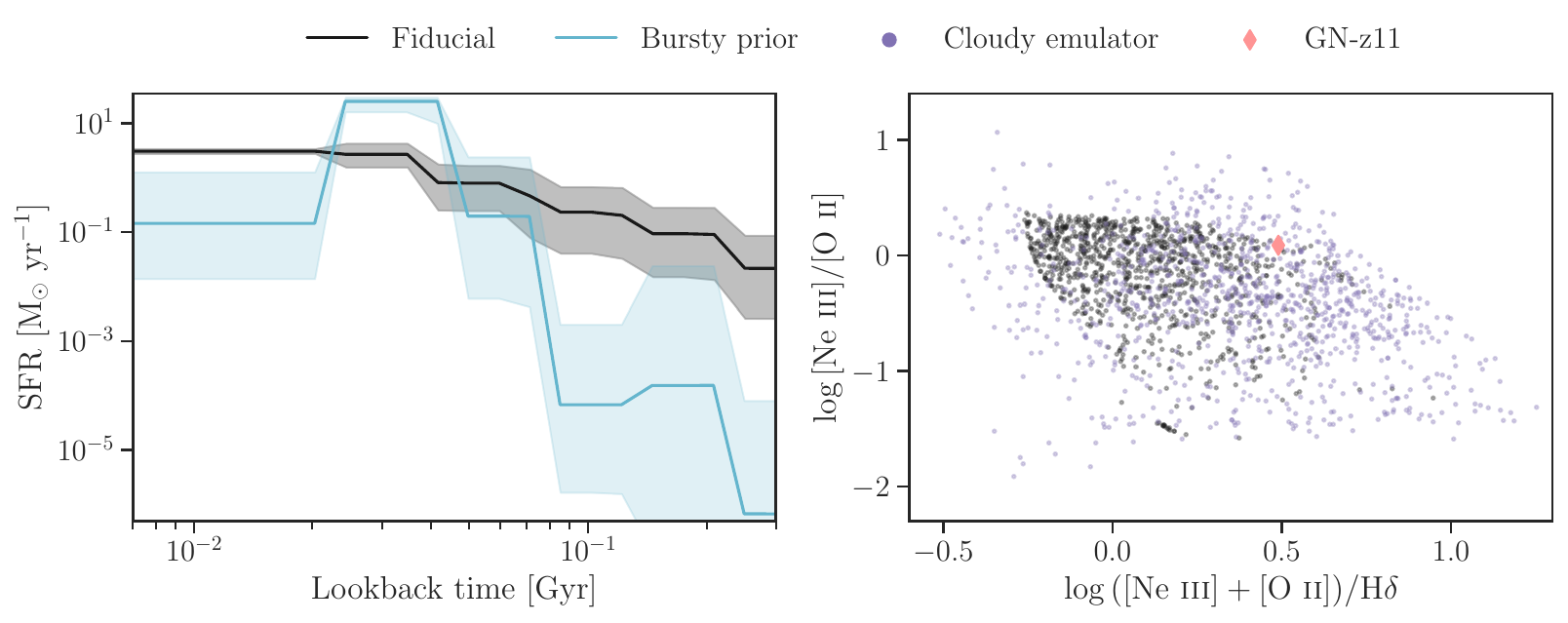}{0.99\textwidth}{}
 }
\caption{Inferred properties for one of the galaxies as an example. (Left) The inferred SFH assuming a rising SFH prior \citep{Wang2023:pbeta} is plotted as a function of lookback time in Gyr in black, whereas that assuming a bursty prior is plotted in cyan. The bursty prior increases stochasticity as expected, and importantly, the SFR is systematically changed by much larger than the size of the formal uncertainties shown in gray shading at almost all lookback times. (Right) Emission line ratios calculated from posterior samples based on the standard \cloudy\ grid \citep{Byler2017} are shown in black, whereas those from the \cloudy\ emulator, \cue, are shown in purple. The measurement of GN-z11 is included for reference \citep{Bunker2023}. \cue\ enables the exploration of a wider parameter space in nebular conditions.
 }
\label{fig:eg_inferred}
\end{figure*}

\subsection{Alternative Initial Mass Functions\label{subsec:res_imf}}

Changing the slope of the IMF induces a systematic shift in the inferred stellar mass and SFR by $\sim 0.3 - 1$ dex, as suggested in Figure~\ref{fig:scatter}. It has a negligible effect on the inferred stellar age. The effects of varying the minimum stellar mass cutoff is shown in the same way in Figure~\ref{fig:scatter_ml}. More specifically, assuming a top-heavy ($x=1.8$) IMF systematically decreases the inferred stellar mass and SFR by $\gtrsim 0.3$ dex; increasing $m_l$ from 0.08 $\msun$ to 0.5 or 1 $\msun$ further enhances the systematic shifts by $\gtrsim 0.1$ dex. Assuming a top-light ($x=2.8$) IMF increases the inferred stellar mass and SFR by $\sim 1$ dex; increasing $m_l$, however, significantly decreases the systematic shifts. With $x=2.8, m_l=1 \msun$, we recover the fiducial Chabrier masses and SFRs. Our results are consistent with \citet{Woodrum2023}, in which different IMFs are used to fit a sub-sample in the JADES photometric catalog \citep{Eisenstein2023,Rieke2023}.

Importantly, the Bayes factors \citep{Trotta2008}, plotted in Figure~\ref{fig:lnz}, indicate that the data are almost completely agnostic to the different forms of IMFs. 
Further intuition can be gained from examining a single fit. This is illustrated in Figure~\ref{fig:eg_modspec}, where the posterior median model photometry and spectra from different IMF cases are plotted. All models can fit the observed photometric data equally well, and the model spectra only differ marginally from each other. The longer wavelength range covered by JWST/MIRI may offer additional distinguishing power. Similar results are found for the rest of the sample.

\subsection{Bursty Star Formation Prior\label{subsec:res_sfh}}

The bursty SFH prior of this paper causes negligible systematic changes in the inferred stellar masses. However, it tends to increase the uncertainty in the inferred SFR averaged over the most recent 10 Myr, SFR$_{10}$, with a subset of the fits having these recent SFRs that significantly deviate from their fiducial values in Figure~\ref{fig:scatter}. In contrast, the inferred SFRs averaged over a longer time scale of the most recent 100 Myr, SFR$_{100}$, are much less affected. Additionally, the bursty prior leads to a systematic decrease in the inferred average stellar ages of all galaxies, of order $0.2-0.5$ dex.

The left panel of Figure~\ref{fig:eg_inferred} shows the inferred SFHs for the object where the most dramatic change in SFR is observed. The increase in the stochasticity in the SFH due to the bursty prior is clearly visible, and at almost all lookback times the star formation rate is systematically changed by much larger than the size of the formal uncertainties.

Similar to the IMF case, diagnostics based on the Bayes factor, as well as the median model photometry and spectra show that the observed photometric data have little power to discriminate between bursty and smooth SFHs in general. This is in spite of the fact that here we contrast extreme opposite scenarios of an extremely bursty prior with a relatively smooth prior. We note that this null result differs from that of a recent study, which claims that stochastic SFHs are found at $z \sim 9$ \citep{Ciesla2023} using the JADES photometric catalog \citep{Eisenstein2023,Rieke2023}. Given the relatively weak constraining power of broad-band photometry coupled with the uncertain redshifts, more informative data, in particular spectroscopic observations, are likely required to reach a definitive conclusion.

\subsection{Flexible Nebular Emission Modeling\label{subsec:res_cue}}

The more flexible nebular emission modeling, enabled by the \cloudy\ emulator, \cue, causes differences in the inferred masses and SFRs that are within $1\sigma$ of the formal uncertainties. However, consistent increases in the masses and SFRs are found of order $\sim 0.2$ dex. 
We note that this mass offset is more significant that an extrapolation from studies at lower redshift (e.g., the $\sim 0.12$ dex difference found in \citealt{Li2022}), which is a natural consequence of the $z>9$ population being more sensitive to nebular modeling than the regular galaxy population.

It is also a curious finding that the scatter in the model comparison plot of Figure~\ref{fig:lnz} is noticeably larger than those of the above two cases---this suggests that, unlike the IMF and SFH timescales, nebular properties may well be able to be constrained from broad-band photometry, and this constraining power would be substantially increased by medium-band photometry.

To gain more insights into the influence of the flexible model, we show the emission line ratios calculated from posterior samples for one of the candidates in Figure~\ref{fig:eg_inferred}. The recent measurement of GN-z11 is included for reference \citep{Bunker2023}. 
Interestingly, the line ratios of GN-z11 seem to occupy a space that is better sampled by \cue. 
We note that this particular galaxy is claimed to host a bright AGN, which also explains its peculiar abundance \citep{Maiolino2023,Scholtz2023}. Line emission due to AGN are capable of being modeled with \cue; however, here we assume the ionizing spectra of our galaxies are dominated by stars, and use \cue\ such that the ionizing input follows the stellar ionizing input. We opt not to allow full freedom in the ionizing spectra since the photometric data lack the constraining power to distinguish between the stellar and AGN contributions. The ability to sample the parameter space occupied by GN-z11 thus indicates that the emulator is flexible enough to model a stellar ionizing spectra mimicking an AGN spectrum.
These results suggest that while more flexible nebular models have a relatively less strong effect on bulk properties inferred from SED-fitting than key uncertainties like the IMF and SFH timescales, they will be very important in interpreting more detailed spectroscopic observations---and they can be observationally constrained. We discuss the various implications from these results in the following section.

\section{Discussion\label{sec:discussion}}

\subsection{Systematic Uncertainties from Fixed Parameters in SED-modeling}

\begin{figure*}
 \gridline{
 \fig{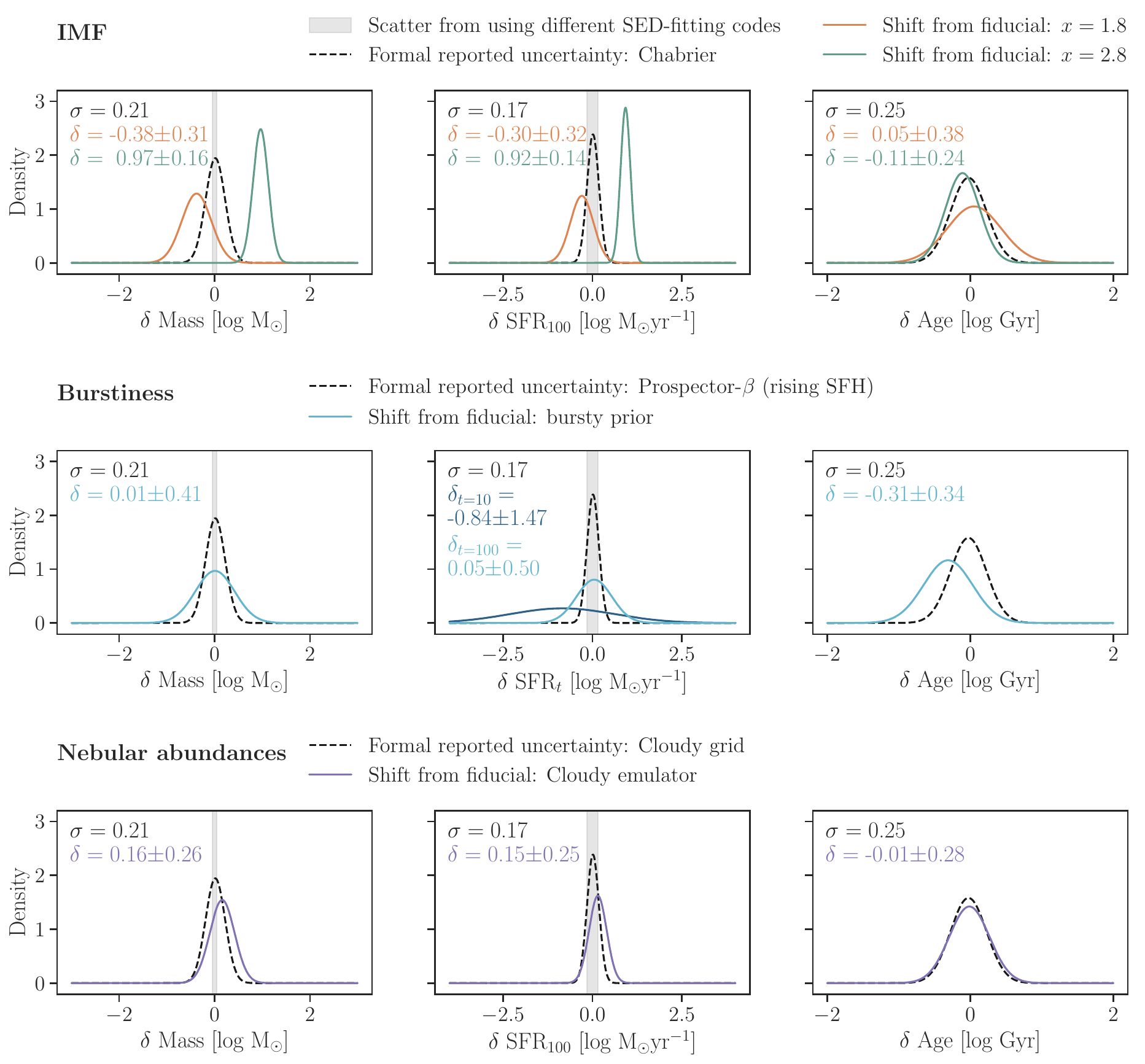}{0.96\textwidth}{}
 }
\caption{Formal reported uncertainties vs. systematics. (Upper panel) The black curve is a Gaussian fit to the stacked posteriors with median subtracted for the Chabrier stellar masses. The fit suggests that the typical reported uncertainty, $\sigma$, is about 0.20 dex. The orange (green) curve is a Gaussian fit to the stacked posteriors of masses assuming a $x=1.8$ ($x=2.8$) IMF, with median of the Chabrier masses subtracted. The mean, -0.38 (0.97), describes the typical systematic shift after changing the IMF, whereas the standard deviation, 0.31 (0.16), indicates the spread in the systematics. Overplotted as the gray shade is the scatter driven by using different SED fitting codes \citep{Pacifici2023}. The other two plots contrast the uncertainties and the systematics in the SFRs and stellar ages, respectively. Critically, the systematics in the inferred stellar mass and SFR due to different IMF choices can be at least comparable to and often larger than the standard uncertainties (posterior moments), and the scatter seen when fitting the an object with different codes. 
(Middle panel) The same as the upper panel, but showing the systematics introduced by adopting a bursty prior instead of a rising SFH prior \citep{Wang2023:pbeta}. The stochasticity in the SFH leads to a large spread in the SFR posteriors. The typical systematic uncertainty in SFR average over the most recent 10 Myr is of order $\sim 0.84$ dex, and with a large spread of $\pm 1.47$ dex. However, SFR averaged over the most recent 100 Myr is less prone to the change in priors. The typical systematic uncertainty decreases to 0.05 dex, with a spread of $\pm 0.50$ dex.
The typical systematic uncertainty in age is $-0.31 \pm 0.34$ dex.
(Lower panel) The typical systematic uncertainty in inferred mass introduced by modeling the nebular physics with more flexibility while retaining consistency with the model stellar populations is comparable ($\sim 0.16$ dex) to the formal uncertainty. A similar systematic uncertainty in SFR ($\sim 0.15$ dex) is also observed.
}
\label{fig:sys_unc}
\end{figure*}

\begin{figure*}
 \gridline{
 \fig{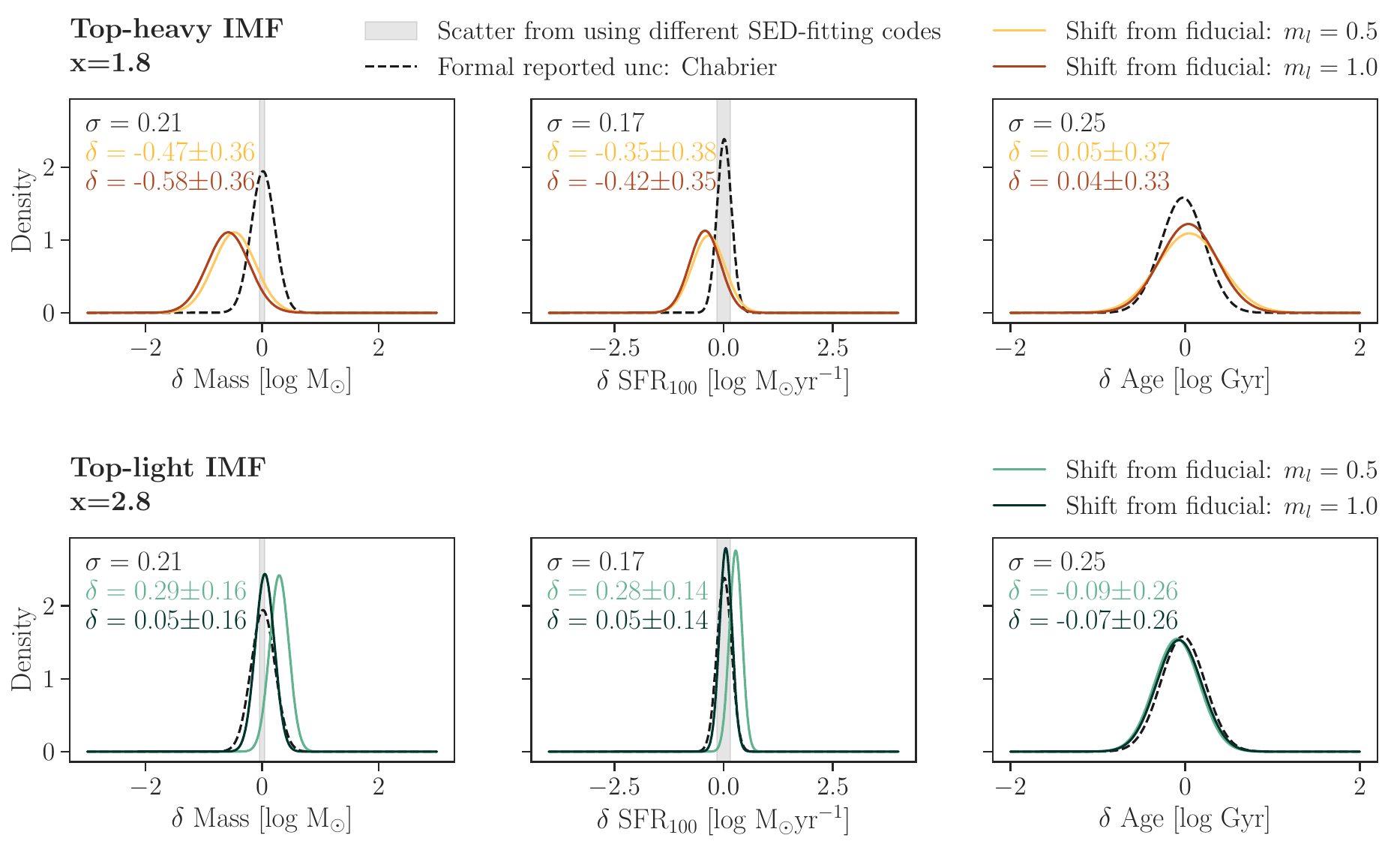}{0.96\textwidth}{}
 }
\caption{Same as Figure~\ref{fig:sys_unc}, but showing the uncertainties for the cases assuming different lower limits, $m_l$, of the IMF. For the top-heavy IMP, increasing $m_l$ leads to larger systematic uncertainties in inferred stellar mass and SFR. For the top-light IMF, increasing $m_l$ leads to smaller systematic shifts owing to less hidden mass.}
\label{fig:sys_unc_ml}
\end{figure*}

Two immediate conclusions can be reached from Section~\ref{sec:res}: first, changing the IMF and the SFH prior produces substantial systematic effects on some of the key stellar population parameters; second, the broad-band photometric observations available from JWST cannot distinguish between the model choices.

It is thus important to ask, given the reasonable model choices and data, what the ``real" uncertainties are for high-redshift galaxies. One way to address this question is by comparing the systematic uncertainties to the reported formal uncertainties. Here a formal uncertainty refers to the 16th--84th percentiles of posterior distributions from a single fit, where the different model choices cannot yet be marginalized. We compare the two kinds of uncertainties as follows.

For each fit, we subtract the medians from the posteriors, which are then stacked to provide an estimate of the average formal uncertainty. The resulting distribution for each of the inferred parameters is well approximated by a Gaussian function. 
We quantify the systematic uncertainty in a similar manner, except that we subtract the median calculated from the fiducial fit from the posteriors. A Gaussian function is fitted to the stacked posteriors, of which the mean corresponds to the average systematic shift, and the standard deviation indicates the range of possible systematic shifts.

The typical formal and systematic uncertainties of stellar mass, SFR, and stellar age are reported in Figure~\ref{fig:sys_unc}. For the case of a non-universal IMF, the usually unaccounted systematic uncertainty in stellar mass ($0.4 \pm 0.3$ dex for a flatter, top-heavy IMF, and $1.0 \pm 0.2$ dex for a steeper, top-light IMF) can be 2--5 times larger than the formal uncertainty ($\sim 0.2$ dex). The same applies to SFR, where the formal uncertainty is found to be $\sim 0.2$ dex, but the systematic uncertainties are $0.3 \pm 0.3$ dex and $1.0 \pm 0.1$ dex for a top-heavy and a top-light IMF, respectively. 
In addition, theoretical models have indicated even more exotic IMF shapes at high redshift (e.g., a flat IMF in \citealt{Chon2022}), or possible environmental dependences \citep{Gunawardhana2011,Jerabkova2018,Yan2019}. Here we only intend to cover the range of the IMF slopes suggested by observations.

Critically, a steep IMF, motivated on both theoretical \citep{Chabrier2014} and observational grounds \citep{Conroy2017,vanDokkum2023}, leads to systematics that always dominate over the error budget from all other uncertainties normally considered. These values are also greater than scatter caused by fitting a sample of galaxies with different SED-fitting codes, which are found to be $\sim 0.1$ dex and $\sim 0.2$ dex for stellar mass and SFR, respectively \citep{Pacifici2023}. This means that a common practice in the literature of assessing uncertainties in SED-fitting processes by comparing multiple codes is substantively underestimating the true uncertainty, as these codes typically use similar assumptions for the IMF, the smoothness of the SFH, and nebular emission.

We note that the minimum and the maximum stellar mass cutoffs, which are held fixed in the above analyses, may further affect the systematic uncertainties. Of particular interest is the low-mass end of the power law, as for a steeper, top-light IMF, the stellar mass would be dominated by stars around the lower limit of the IMF. Thus far, we focus on the discussion on the varying IMF slopes because low-mass stars make up a negligible fraction of the total light. We additionally include Figure~\ref{fig:sys_unc_ml} showing the systematic uncertainties with increasing lower mass limits to illustrate the degeneracy between $m_l$ and inferred stellar mass.
For the top-light IMF, assuming $m_l= 0.5 \msun$, the systematic uncertainties of stellar mass and SFR decrease to $0.3 \pm 0.2$ dex, comparable to the formal uncertainties; assuming $m_l=1 \msun$, the inferred parameters exhibit negligible differences comparing to the fiducial Chabrier values. 
It is, however, also easy to imagine performing a similar analysis for IMFs with increasing hidden mass, either owing to lower $m_l$ or steeper slopes at the low-mass end. These alternatives are difficulty to be ruled out as low-mass stars would be largely invisible outside the Milky Way, and would increase the systematic uncertainties.
For the top-heavy IMF, the results are less sensitive to the lower limit. The stellar masses are systematically decreased by $0.5 \pm 0.4$ ($0.6 \pm 0.4$) dex, and the SFRs are similarly decreased by $0.4 \pm 0.4$ dex, if $m_l=0.5\msun$ ($1\msun$).

The same analysis is performed for the SFH priors which assume different timescales for changes. 
Also plotted in Figure~\ref{fig:sys_unc}, we find that stochastic SFHs, which are expected at high redshift \citep{Tacchella2016,Faucher-Gigueere2018,Legrand2022,Dome2023}, produce significant systematics in SFR averaged over the most recent 10 Myr ($0.8 \pm 1.5$ dex), and in stellar age ($0.3 \pm 0.3$ dex). The formal uncertainties in these two parameters are $\sim 0.2$ dex and $\sim 0.3$ dex, respectively. It is worth noting that the systematics in the SFR can be mitigated by averaging the SFR over a longer time scale. Using the most recent 100 Myr, the systematic uncertainty in SFR decreases to $0.1 \pm 0.5$ dex.
In addition, it may be surprising to see that the bursty prior does not lead to significant systematics in the inferred stellar mass ($0.0 \pm 0.4$ dex). This is likely because of the construction of our fiducial SFH prior from \citet{Wang2023:pbeta}, which has been shown to recover stellar mass reasonably well even in the presence of very bursty SFHs from simulations \citep{Narayanan2023}.

In contrast to the substantial systematic shifts found in the above two cases, a more flexible nebular emission model results in systematic uncertainties in mass and SFR ($0.2 \pm 0.3$ dex) that are comparable to the formal uncertainty ($\sim 0.2$ dex). The smaller systematic changes mean that even in the case of very high emission line equivalent widths expected in the first galaxies, the assumption of solar-scaled nebular abundance patterns, which are not at all expected to hold at early times, do not cause factor-of-two or more systematics in the inferred masses, SFRs, and ages. Furthermore, the larger scatter in the Bayesian evidences and the differences in the observed line ratios suggest that this is a solvable systematic with future observations---spectroscopy, or even photometry with relatively higher spectral resolution.

A counterintuitive finding, however, is that the estimates for stellar mass and SFR of one galaxy, ID 14088, using the model with increased flexibility are much stronger constrained than the original model. In addition, the Bayesian evidence strongly favors the original model for this particular galaxy. Upon examination, the fiducial fit has a bimodal posterior distribution for the stellar-phase metallicity, but one of the solutions is disfavored by \cue, thus decreasing the error bars. A possible reason is that while \cue\ finds a marginally better single mode with the increased flexibility, the likelihood of this mode is not high enough to compensate for the added four parameters.

Finally, we note that the contribution from AGN can have significant influence on the stellar mass estimates \citep{DSilva2023}. In this work, we only model the mid-infrared emission from the AGN, and find that our sample has low AGN contribution. The mean AGN to galaxy bolometric luminosity is $\sim 0.004$. However, we caution that the wavelength coverage of our photometric data is not expected to process constraining power for the particular model considered. The AGN modeling in \prospector\ will be improved in future works.

\subsection{Implications for Early Galaxy Formation and Possible Ways Forward}

Given the significant systematics in parameters inferred by SED-fitting combined with the lack of short-term prospects for observational constraints on said parameters, one could argue for including these systematic uncertainties in our error budget for all galaxies. 
Indeed, the IMF-induced systematics are often larger than the uncertainties and systematics usually considered in the stellar mass functions \citep{Behroozi2010,Conroy2013,Speagle2014,Grazian2015,Santini2015,Carnall2018,Leja2019:3dhst,Furtak2021}. 

More recently, various papers have challenged the validity of galaxy formation models or $\Lambda$CDM in light of newly discovered galaxy populations with JWST (e.g., \citealt{Casey2023,Labbe2023,Xiao2023}). A non-universal IMF can serve as a possible explanation for the overly massive galaxies, as pointed out in \citet{Boylan-Kolchin2023,Steinhardt2023}.
Taking into account the systematic uncertainties found in this paper can, for example, bring the star formation efficiency expected from the \citet{Labbe2023} sample into $1 \sigma$ consistency with a reasonable value of $\lesssim 0.32$ at $z > 8$, and that expected from the \citet{Xiao2023} sample to $\lesssim 0.2$ at $z > 5$; thus removing the need for exotic cosmological models (e.g., \citealt{Gong2023,Padmanabhan2023,Parashari2023}). This is intriguing given that the modifications in the power spectrum necessary for solving the tension with $\Lambda$CDM can induce conflicts at lower redshifts \citep{Sabti2023}.

Unfortunately, observationally constraining the IMF beyond the Milky Way is a known challenge. Thus far most of the meaningful constraints come from independent measurements on the mass-to-light ratio from gravitational lensing \citep{Treu2010,vanDokkum2023} or dynamical modeling \citep{Cappellari2013,Posacki2015}, or from direct observational signatures of low-mass stars in high S/N, $R\sim3000$ near-IR spectra \citep{Conroy2012,vanDokkum2017}. These approaches are difficult to generalize to a statistical sample at cosmological distance. 

Therefore, it seems necessary to fully incorporate the systematic uncertainties in the error budget. This is, however, non-trivial via conventional techniques. To start, including IMF as a free parameter requires rebuilding the stellar library in every fit, which makes it infeasible to sample the likelihood surface within a reasonable amount of time. 
Possible solutions include machine-learning-accelerated SED fitting of individual galaxies \citep{Hahn2022,Wang2023:sbi}, or population-level inference \citep{Alsing2023,Li2023}. For the former, once a suitable training set is constructed, it becomes straightforward to marginalize over IMF variations. The latter approach directly makes the population distribution the inference objective, which drastically increases the computational efficiency.

The over-abundance of high-redshift galaxies is another interesting discovery from early JWST observations (e.g., \citealt{Yung2023}). Bursty star formation has been invoked as a possible explanation \citep{Pallottini2023,Shen2023,Sun2023}. While we find that adapting a bursty prior can fit the photometric observations equally well, the current data do not offer evidence for or against such expectation. Understanding the prevalence of bursty star formation, and quantifying the resulting selection effects (i.e., galaxies oscillating in and out of the observable sample due to SFR variability) likely requires at least spectroscopy in order to measure the different SFR and SFH indicators that can constrain burstiness. Perhaps the most convincing evidence would stem from exploring population distributions of these spectral indicators.

As for nebular physics, the \cloudy\ emulator, \cue, allows for the exploration of a wide parameter space while retaining consistency with stellar populations; particularly relevant to this work is the BPT region occupied by GN-z11, which is sparsely sampled in the standard \cloudy\ grid \citep{Byler2017}. This indicates that, with more informative spectroscopic or medium-band photometric data, the flexible model is a promising approach to characterize the exotic nebular conditions in the early universe.

Having discussed the systematic effects on the inferred galaxies properties, we would like to add that none of the alternative models leads to dramatic changes in the inferred photometric redshifts. For the $z>9$ range studied in this work, the photometric redshifts are almost entirely determined by the Lyman break. 
The \cue\ parameterization does not influence the broad-band photometry substantial enough to change the results.
This means that the uncertainties in the photometric redshifts are mainly driven by the modeling of Lyman-$\alpha$ and its interaction with the intergalactic medium, although nebular emission lines such as H$\alpha$ have been shown to be able to resemble the Lyman break in photometric data by spectroscopic follow-ups \citep{2023arXiv230315431A}. In the near future, we plan to incorporate a phenomenological model to describe the radiative transfer process (e.g., damping wings; \citealt{Curtis-lake2023}). 
With that being said, the transfer of resonant lines in gas is an intricate process, the accurate description of which requires numerical radiative transfer simulations (e.g., \citealt{Gronke2017,Michel-Dansac2020}). There is much room for improvement to fully capture the complexity of the propagation of Lyman-$\alpha$.

Finally, we note that the systematics analyzed for the $\zphot > 9$ sample of this paper are expected to be present across redshifts, albeit with varying degrees of importance. On the observational front, JWST has been revealing new galaxy populations and establishing statistical samples, reaching uncharted regions in the cosmic history. Sophisticated modeling of galaxy populations that encapsulates the full uncertainties, analogous to the framework proposed for redshift distribution inference in \citet{Alsing2023}, would be invaluable in forming a coherent narrative of galaxy evolution over the observed dynamic range.

\section{Conclusions\label{sec:conclusion}}

In this paper, we quantify the systematic uncertainties from different model choices on inferred high-redshift galaxy properties. We conduct our experiment on a $z>9$ sample selected from the publicly available UNCOVER SPS catalog \citep{Wang2023:uncover}. As a large-scale application of the \prospector\ Bayesian inference framework, where redshifts and stellar population parameters are inferred jointly, the UNCOVER SPS catalog provides a baseline against which the fundamental assumptions in SED modeling can be tested. In particular, we investigate the three model choices as follows.

First, we fit the photometry assuming a flatter/steeper IMF in addition to the fiducial Chabrier IMF. The photometric data are firmly unable to distinguish between particular forms of IMF. 
We find that a flatter IMF ($x=1.8$) tends to systematically decrease the inferred stellar mass by $\sim 0.4$ dex, and SFR by $\sim 0.3$ dex. A steeper ($x=2.8$) IMF typically increase the inferred stellar mass and SFR both by $\sim 1.0$ dex. These values are notably larger than the reported uncertainty of $\sim 0.2$ dex on mass and SFR from fits assuming a fixed IMF, and also the scatter found when fitting data using different SED codes ($\sim 0.1$ dex in mass and $\sim 0.2$ dex in SFR; \citealt{Pacifici2023}). However, these results can be sensitive to the lower mass limits of the IMF, which increase or decrease the systematic uncertainties depending on the change in the amount of hidden mass. Observationally, this is hard to measure, since the total integrated light is dominated by massive stars.
Nevertheless, taken together with the lack of IMF constraints at high redshift and the motivations for a non-universal IMF on observational as well as theoretical grounds, we infer that the usually reported uncertainties on stellar mass and SFR are likely to be underestimated.

Second, we adopt an extremely bursty star formation prior, in contrast to the fiducial rising SFH prior proposed in \citet{Wang2023:pbeta}. Similar to the varying IMF case, the Bayes factor shows no particular preference given the photometric data. 
The bursty prior leads to $> 1$ dex deviations from the fiducial SFRs averaged over the most recent 10 Myr in some cases, however the large systematics can be mitigated by averaging the SFRs over longer timescales, e.g., 100 Myr. The typical systematic decrease is $\sim 0.8$ dex in inferred SFR$_{10}$, which is greater than the formal uncertainty of $\sim 0.2$ dex, and $\sim 0.1$ dex in SFR$_{100}$. Therefore, care must be taken when attempting to infer burstiness by comparing SFRs averaged over different time scales. Additionally, the bursty prior results in marginally younger stellar ages for the full population, of an order of $\sim 0.3$ dex, comparable to the formal uncertainty.
It is worth emphasizing that, while being able to fit the observations equally well, the inferred SFR assuming the bursty prior is systematically changed by much larger than the size of the formal uncertainties at almost all lookback times. This means that the inference on SFH from broad-band photometry is prior-dominated rather than likelihood-dominated; in other words, choosing the appropriate prior is critical for SFH studies, as pointed out in \citet{Leja2019:nonpar,Suess2022,Tacchella2022:metal}.

Third, we implement a \cloudy\ emulator (\cue; Li et al. in prep.) into \prospector, which allows us to model the exotic nebular conditions in the early universe while retaining consistency with the assumed stellar populations. This results in systematic uncertainties in mass and SFR ($\sim 0.2$ dex) that are comparable to the formal uncertainty. Interestingly, utilizing \cue\ also leads to better explored parameter space on the BPT diagram where GN-z11 resides. This implies that flexible nebular emission modeling would be valuable for both for describing and interpreting chemical evolution given more informative (higher resolution) data.

To conclude, the new redshift frontier established by JWST pushes stellar population models into new, exciting, and largely uncalibrated regimes. This work represents a first step toward a comprehensive error budget in inferred galaxy properties from SED modeling, paving the way for an accurate accounting of the intricate processes governing galaxy formation and evolution.

\section*{Acknowledgments}

We thank the anonymous referee for the helpful comments. We are also grateful to Ben Johnson for his help in developing the \cloudy\ emulator.
This work was performed in part at Aspen Center for Physics, which is supported by National Science Foundation grant PHY-2210452. 
B.W. and J.L. acknowledge support from JWST-GO-02561.022-A.
H.A. is supported by CNES, focused on the JWST mission. H.A. also acknowledges support from the Programme National Cosmology and Galaxies (PNCG) of CNRS/INSU with INP and IN2P3, co-funded by CEA and CNES. P.D. acknowledges support from the Dutch Research Council (NWO) through the award of the VIDI Grant 016.VIDI.189.162 (``ODIN") and the European Commission's and University of Groningen's CO-FUND Rosalind Franklin program.
L.J.F acknowledges support by Grant No. 2020750 from the United States-Israel Binational Science Foundation (BSF) and Grant No. 2109066 from the United States National Science Foundation (NSF).
J.E.G. acknowledges NSF/AAG grant \#1007094.
The work of C.C.W. is supported by NOIRLab, which is managed by the Association of Universities for Research in Astronomy, Inc. (AURA) under a cooperative agreement with the NSF.

This work is based in part on observations made with the NASA/ESA/CSA James Webb Space Telescope. The data were obtained from the Mikulski Archive for Space Telescopes at the Space Telescope Science Institute, which is operated by AURA, under NASA contract NAS 5-03127 for JWST. These observations are associated with JWST-GO-2561, JWST-ERS-1324, and JWST-DD-2756. Support for program JWST-GO-2561 was provided by NASA through a grant from the Space Telescope Science Institute under NASA contract NAS 5-26555. This research is also based on observations made with the NASA/ESA Hubble Space Telescope obtained from the Space Telescope Science Institute under NASA contract NAS 5–26555. These observations are associated with programs HST-GO-11689, HST-GO-13386, HST-GO/DD-13495, HST-GO-13389, HST-GO-15117, and HST-GO/DD-17231.
The specific observations analyzed can be accessed via \dataset[10.17909/nftp-e621]{http://dx.doi.org/10.17909/nftp-e621}, and the SPS catalog corresponds to the following copy deposited to zenodo:
\dataset[10.5281/zenodo.10223792]{https://doi.org/10.5281/zenodo.10223792}.

Computations for this research were performed on the Pennsylvania State University's Institute for Computational and Data Sciences' Roar supercomputer. This publication made use of the NASA Astrophysical Data System for bibliographic information.

\facilities{HST(ACS,WFC3), JWST(NIRCam, NIRSpec)}
\software{Astropy \citep{2013A&A...558A..33A, 2018AJ....156..123A,2022ApJ...935..167A}, Dynesty \citep{Speagle2020}, Matplotlib \citep{2007CSE.....9...90H}, NumPy \citep{2020Natur.585..357H}, Prospector \citep{Johnson2021}, SciPy \citep{2020NatMe..17..261V}}

\newpage
\appendix

\begin{figure*}
 \gridline{
 \fig{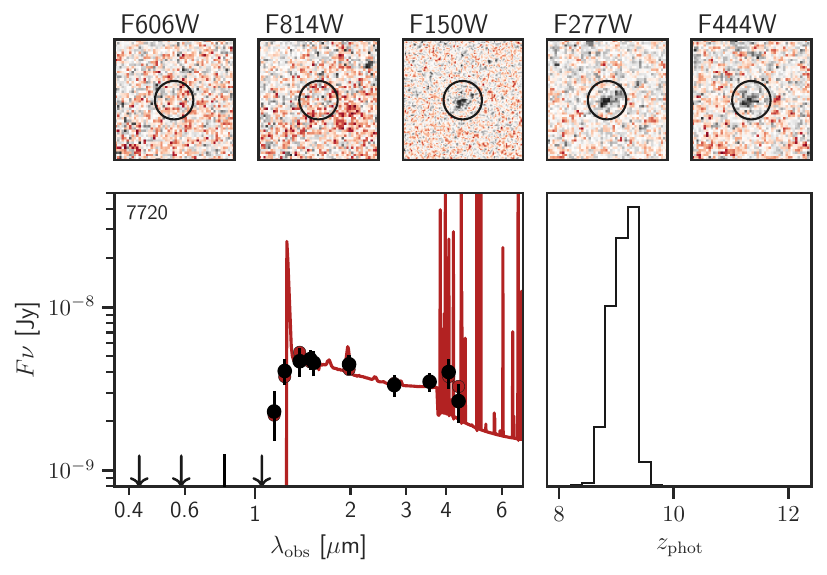}{0.45\textwidth}{}
 \fig{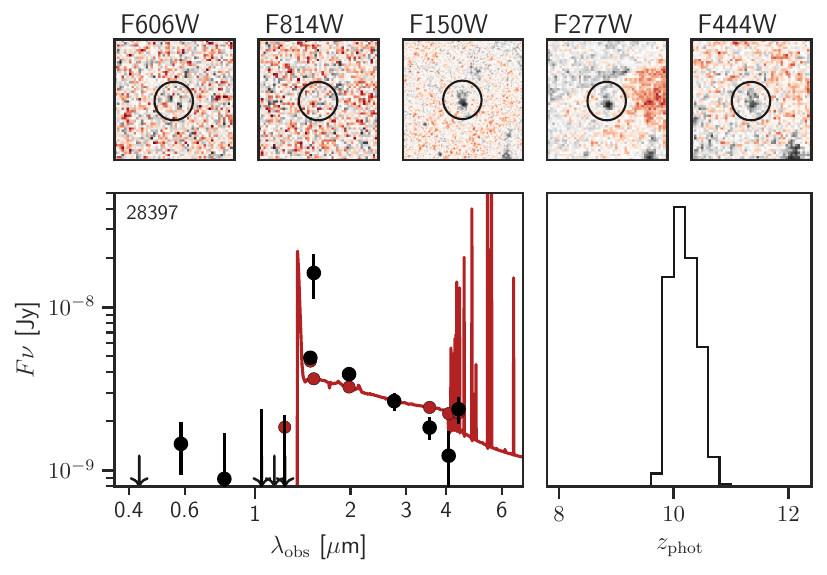}{0.45\textwidth}{}
 }
 \gridline{
 \fig{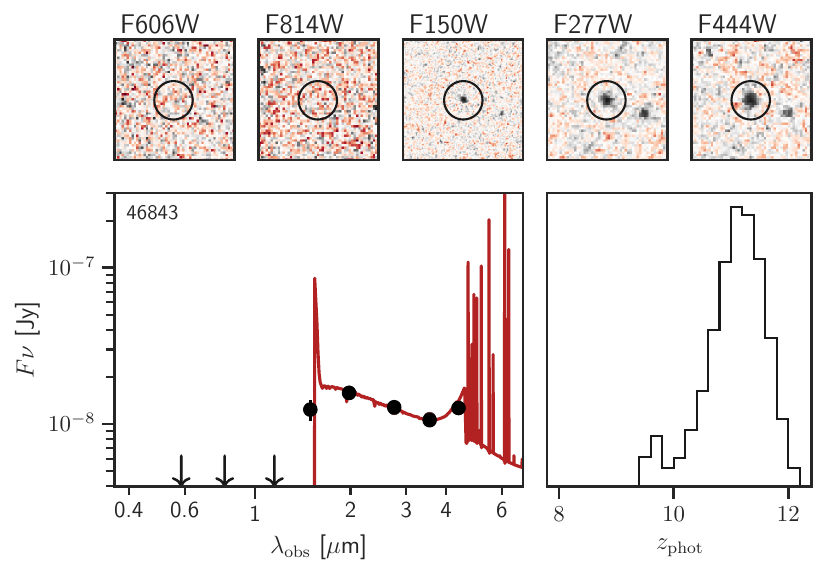}{0.45\textwidth}{}
 \fig{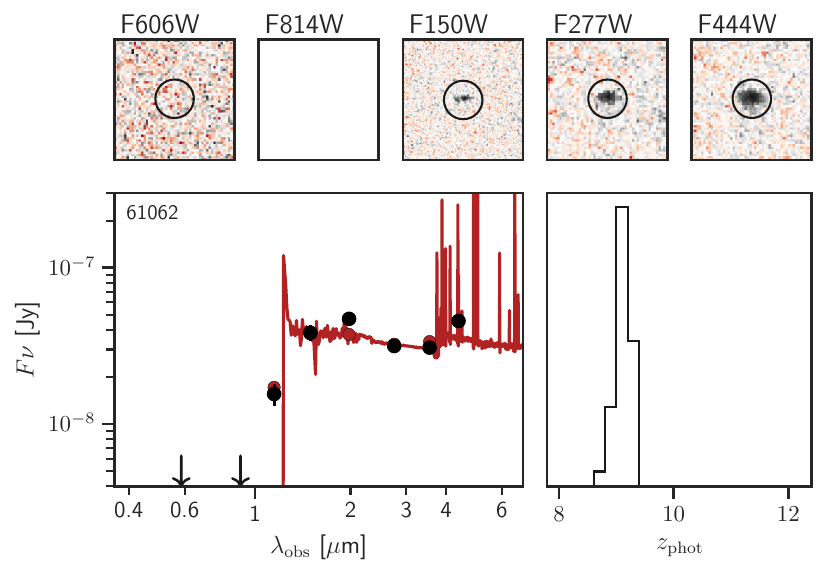}{0.45\textwidth}{}
 }
\caption{New candidates in this paper. For each candidate, we shown cutouts including two HST bands (F606W, and F814W) and three JWST bands (F150W, F277W, and F444W), observed photometry in black, best-fit model in red, and the marginalized probability distribution of redshift posteriors. The black circle on each cutout indicates an aperture size of 0.32\arcsec{}, within which the photometry is extracted.}
\label{fig:new}
\end{figure*}

\begin{figure*}
 \gridline{
 \fig{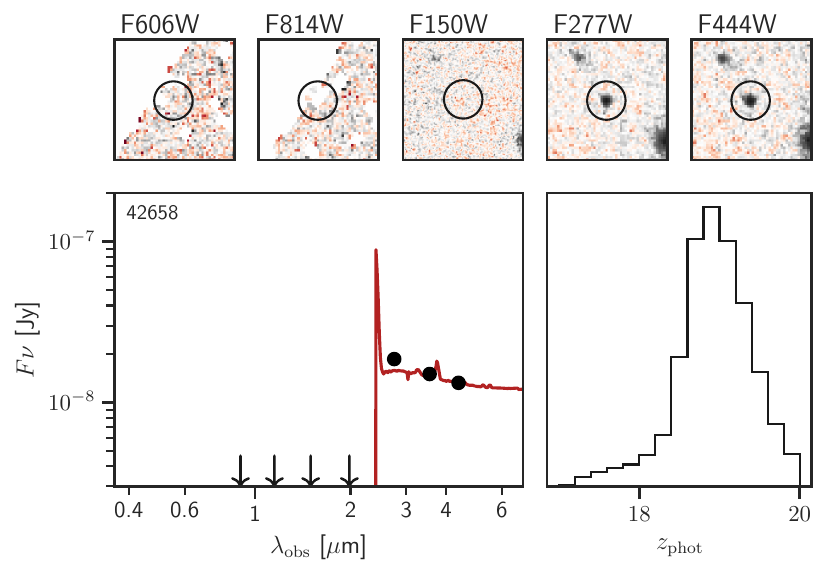}{0.45\textwidth}{}
 }
\caption{A possible artifact. ID 42658 only appears in the first epoch, and falls in the detector gap in the second epoch.}
\label{fig:artifact}
\end{figure*}

We supplement further details on the $z>9$ sample of this paper in this appendix. Figure~\ref{fig:new} shows the new candidates, whereas Figure~\ref{fig:artifact} shows a possible data artifact. 
Additionally, we list the three objects with $\zspec > 9$, but are not included in the sample of this paper in Table~\ref{tab:specz}. ID 11701 only have data in 6 NIRCam filters, and ID 21347 is identified as a broad-line AGN \citep{Kokorev2023}. Both sources show uncommon spectral signatures, and it is thus understandable that our photometric redshifts are inaccurate. ID 31955 has a likely spurious detection in a dropout band, albeit with large uncertainty, which skews the posterior mass in redshift to lower values.

\begin{deluxetable*}{lccccccp{0.4\linewidth}}
\tablecaption{Discrepant Photometric and Spectroscopic Redshifts\label{tab:specz}}
\tablehead{
\colhead{ID} & \colhead{$\rm ID_{MSA}$} & \colhead{RA} & \colhead{Dec} & \colhead{$z_{\rm spec}$\tablenotemark{\scriptsize{a}}} & \colhead{$z_{\rm phot, ml}$} & \colhead{$z_{\rm phot, med}$}  & \colhead{Notes}
}
\startdata
11701 & 10646 & 3.63696 & -30.40636 & 8.510 & 0.275 & $0.272^{+0.012}_{-0.012}$ & Only have data available in 6 JWST filter bands; will be presented in Weaver et al., in prep.\\ 
21347 & 20466 & 3.64041 & -30.38644 & 8.500 & 1.305 & $1.309^{+0.132}_{-0.098}$ & Broad-line AGN \citep{Kokorev2023} \\
31955 & 31028 & 3.54417 & -30.37032 & 9.740 & 0.522 & $0.567^{+0.132}_{-0.094}$ & Noisy photometric measurement in F814W\\ 
\enddata
\tablenotetext{a}{NIRSpec/Prism confirmation conducted during the second phase of the UNCOVER survey (\citealt{Fujimoto2023:uncover}; Price et al. in prep).}
\end{deluxetable*}

\bibliography{uncover_systematics_wang.bib}

\end{document}